\documentclass[mypaper,8pt,twoside]{CoAst}
\usepackage{epsf,graphicx,fancyhdr}
\renewcommand{\chaptermark}[1]{\markboth{#1}{}}

\renewcommand{\headrulewidth}{0pt}
\newcommand{\teff}{\ensuremath{T_{eff}}}             
\newcommand{\logg}{\ensuremath{\log g}}                     

\def\cd {d$^{-1}$}

\newcommand{\kopf}{\small\itshape Comm. in Asteroseismology\\ Vol. 144, 2003}
\newcommand{\Authors}[1]{\begin{center}\normalsize\bf\sf #1 \end{center}}

\renewcommand{\author}[1]{\begin{center}\normalsize\bf\sf #1 \end{center}}
\newcommand{\Address}[1]{\begin{center}\small\sf #1 \end{center}}

\renewenvironment{abstract}{\section*{Abstract}\normalsize\sf}{}
\newcommand{\References}[1]{\begin{flushleft}{\large References\\}\vspace*{2mm}\small #1 \end{flushleft}}

\newcommand{\chapterDSSN}[2]{\chapter[\sf\normalsize #1\\ \footnotesize \hspace*{5mm}by #2 \sf\normalsize][]{#1\\}\rhead[\fancyplain{}{\sf\footnotesize \center{#1}}]{\fancyplain{}{\sffamily\thepage}}\lhead[\fancyplain{\kopf}{\sffamily\thepage}]{\fancyplain{\kopf}{\sf\footnotesize \center{#2}}}}

\renewcommand{\chaptername}{}
\newcommand{\acknowledgments}[1]{\vspace*{5mm}\noindent\begin{bf}Acknowledgments. \end{bf} #1}

\begin{document}
\newcommand{\nua}{$\nu _1$}
\newcommand{\nub}{$\nu _2$}
\newcommand{\nuc}{$\nu _3$}
\newcommand{\nud}{$\nu _4$}
\newcommand{\nurot}{$\nu _{\rm rot}$}
\newcommand{\nuorb}{$\nu _{\rm orb}$}

\sf

\chapterDSSN{Analysis of {\sc mercator} data Part I: variable B stars}{P. De Cat,
  M. Briquet, C. Aerts et al.}

\setcounter{footnote}{5}

\Authors{P. De Cat$^{1,2}$, M. Briquet$^{2,}$\footnote{Postdoctoral Fellow of
the Fund for Scientific Research, Flanders}, C. Aerts$^{2,3}$,
K. Goossens$^2$, S. Saesen$^2$, J. Cuypers$^1$, K. Yakut$^2$,
R. Scuflaire$^4$, M.-A. Dupret$^5$ and many observers}  
\Address{$^1$ Royal Observatory of Belgium, B-1180 Brussel, Belgium\\
$^2$ Instituut voor Sterrenkunde, Katholieke Universiteit Leuven, B-3001 Leuven, Belgium\\
$^3$ Department of Astrophysics, Radboud University Nijmegen, 6500 GL Nijmegen, the Netherlands\\
$^4$ Institut d'Astrophysique et de G\'eophysique, Universit\'e de Li\`ege, B-4000 Li\`ege, Belgium\\
$^5$ Observatoire de Paris, LESIA, 92195 Meudon, France}

\noindent
\begin{abstract}
We re-classified 31 variable B stars which were observed more than 50 times in
the Geneva photometric system with the {\sc p7} photometer attached to the
{\sc mercator} telescope (La Palma) during its first 3 years of scientific
observations. 
HD\,89688 is a possible $\beta$\,Cephei/slowly pulsating B star hybrid and the
main mode of the {\sc corot} target HD\,180642 shows non-linear effects.
The Maia candidates are re-classified as either ellipsoidal variables or
spotted stars.  
Although the mode identification is still ongoing, all the well-identified
modes so far have $\ell \le 2$.
\end{abstract}

\section{Introduction}

\begin{figure}
\begin{center}
\resizebox{0.99\textwidth}{!}{\rotatebox{270}{\includegraphics{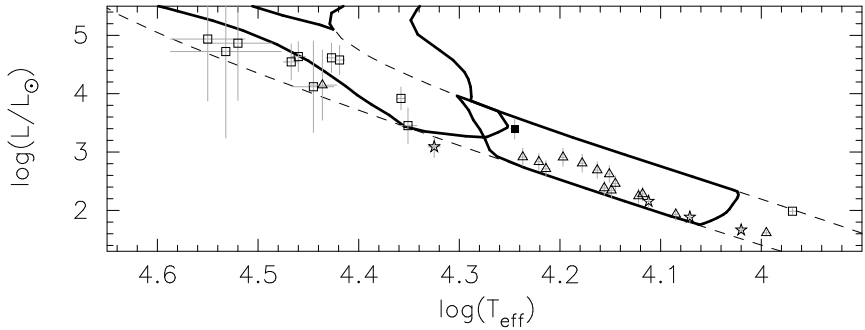}}}
\end{center}
\caption{\label{hr} Position in the H-R diagram of the 31 variable B stars discussed in this
paper.
The candidate $\beta$\,Cephei stars, slowly pulsating B stars and Maia stars
are respectively given with squares, triangles and stars.
HD\,89688 is given with a full symbol.
The ZAMS and TAMS are given with dashed lines, and the theoretical instability
strips for $\beta$\,Cephei and slowly pulsating B stars, as given by Pamyatnykh
(1999), with full lines.
}
\end{figure}

The {\sc mercator} telescope is a 1.2-m telescope located on the Roque de los
Muchachos observatory on La Palma (Spain). 
Since the start of scientific observations in spring 2001, this telescope has
been intensively used to observe variable B, A and F main sequence stars with
the {\sc p7} photometer, providing quasi-simultaneous observations in the 7
passbands of the Geneva photometric system.  
The first results obtained after 18 months of observations were already
presented by De Cat et al. (2004) and De Ridder et al. (2004).
We now present results after 3 years of collecting data.
In Part I (this paper), the analysis of the 9023 datapoints of variable B
stars is discussed while Paper II (Cuypers et al., these proceedings) focuses
on the analysis of the 5149 datapoints of variable A and F stars.

We here restrict ourselves to the 31 variable B stars which were not included
in multi-site campaigns and which were observed at more than 50 epochs. 
Based on the photometric observations gathered with the satellite mission {\sc
hipparcos}, these objects were previously classified as candidate
$\beta$\,Cephei stars ($\beta$\,Ceps; main-sequence B\,0--3 stars pulsating in
low order, low degree $p/g$-modes with periods of 3--8~h), slowly pulsating B
stars (SPBs; main-sequence B\,3--B\,9 stars pulsating in high order, low
degree $g$-modes with periods of 0.5--3~d) and Maia stars (Maias; variable
main-sequence stars situated between the SPBs and the $\delta$\,Scuti stars).
They are respectively given with squares, triangles and stars in
Fig.\,\ref{hr}.

\section{Frequency analysis} \label{fa}

The time series in the Geneva passbands and colours were both subjected to a
detailed frequency analysis with the PDM (Stellingwerf 1978) and Lomb-Scargle
(Scargle 1982) methods. 
Since our ground-based data-sets suffer from strong aliasing, the space-based
photometric observations of the {\sc hipparcos} satellite proved to be very
useful to extract the physical frequencies in some cases.
Our results enable us to re-classify the stars into the following categories
by using the same criteria as De Cat et al. (2004):\\
$\bullet$ SPBs: 11 multiperiodic (HD\,1976, 3379, 21071, 25558, 28114, 28475, 179588,
182255, 191295, 206540, 222555), and 2 monoperidic (HD\,138003, 208057)\\  
$\bullet$ $\beta$\,Ceps: 6 multiperiodic (HD\,13745, 13831, 14053, 21803, 180642, 203664)\\ 
$\bullet$ Hybrid star: HD\,89688\\
$\bullet$ Spotted stars: HD\,46005, 154689, 169820\\ 
$\bullet$ Ellipsoidal variables: HD\,24094, 112396, 149881, 208727\\ 
$\bullet$ Be star: HD\,180968\\
$\bullet$ Constant stars: HD\,19374, 214680, 217782\\
For all the periodograms and phase diagrams, we refer to De Cat et al. (in
preparation). 
For HD\,89688, we now have {\it marginal} evidence for the SPB-like frequency
0.7965(6)~\cd\ (or one of its aliases), while the {\sc hipparcos} photometry
points towards $\beta$\,Cep-like frequency 7.3902(5)~\cd.  
Its position in the H-R diagram is compatible with the classification as a
hybrid star (full symbol in Fig.\,\ref{hr}).
For the multiperiodic {\sc corot} target HD\,180642, the first mode is a high
amplitude mode which shows non-linear effects. 
We detect up to the second harmonic of \nua\,= 5.486971(6)~\cd, making it only
the second $\beta$\,Cep star for which more than one harmonic is observed
(Aerts et al., in preparation).
Note that for all the Maias, i.e. HD\,46005, 154689, 169820,
208727, the observed variations can be explained by mechanisms other than
pulsations. 

\section{Mode identification} \label{mi}

\begin{figure*}
\begin{center}
\resizebox{0.99\textwidth}{!}{\rotatebox{270}{\includegraphics{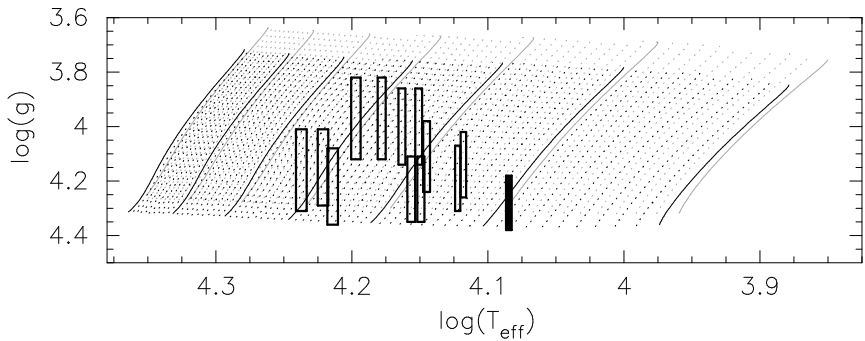}}}
\end{center}
\caption{\label{hrmodel} Presentation of the grids of equilibrium models used
for the mode identification of the SPBs.
The evolution tracks of grid 1 and 2 (see text) with masses between
2--8~$M_{\odot}$ in steps of 0.1~$M_{\odot}$ are respectively given in light
and dark grey.
The boxes represent the observed errorboxes of the 13 SPBs in our sample.
The filled box corresponds to HD\,179588.
}
\end{figure*}

\begin{figure*}
\begin{center}
\resizebox{0.49\textwidth}{!}{\rotatebox{270}{\includegraphics{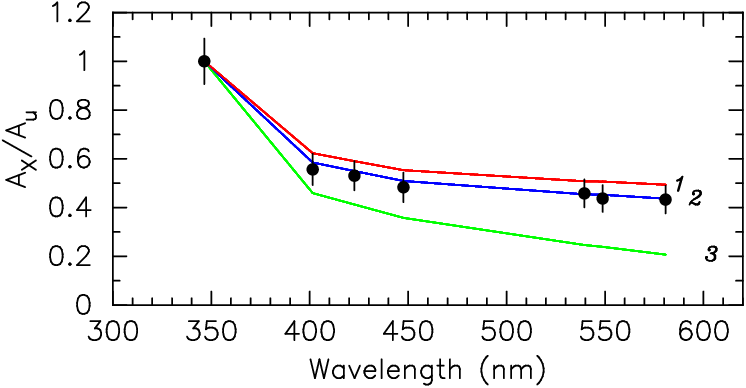}}}
\resizebox{0.49\textwidth}{!}{\rotatebox{270}{\includegraphics{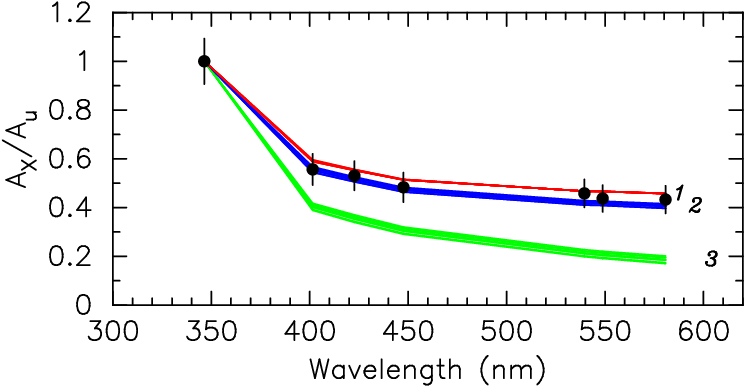}}}\\
\resizebox{0.49\textwidth}{!}{\rotatebox{270}{\includegraphics{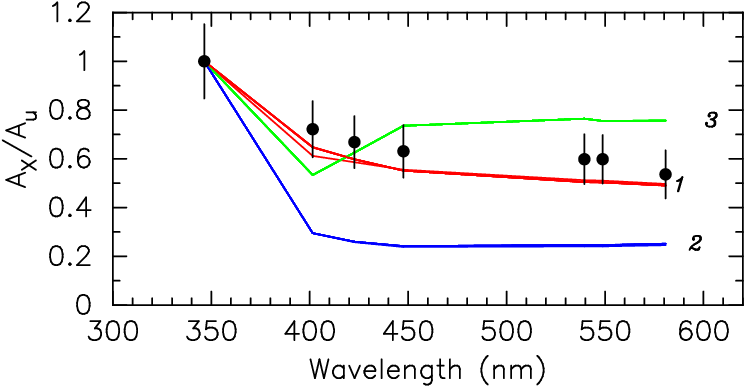}}}
\resizebox{0.49\textwidth}{!}{\rotatebox{270}{\includegraphics{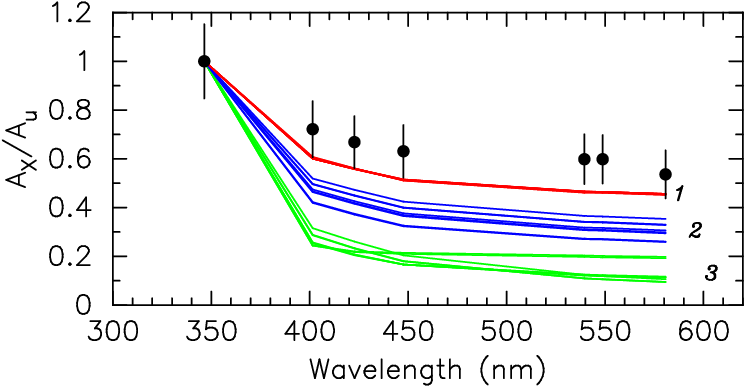}}}
\end{center}
\caption{\label{mifig} Photometric mode identification for
\nua\,=  0.856543(15)~\cd\ (top) and \nub\,= 2.04263(5)~\cd\ (bottom) of
HD\,179588. 
For each theoretical model within the observed range of $\log$(\teff) and
\logg, the theoretical amplitude ratios for modes with $\ell$\,= 1, 2 and 3
are represented with an individual line.  
The dots indicate the observed amplitude ratios and their standard error.
The left and right panels show the results obtained with grid 1 and 2
respectively (see text).
}
\end{figure*}

For the mode identification, we applied the method of the photometric
amplitudes (Dupret et al. 2003) in which the observed and theoretical
amplitude ratios relative to the amplitude in the $U$ filter are compared. 
For the SPBs, we confronted the results based on 2 grids of equilibrium models
(Fig.\,\ref{hrmodel}). 
Grid 1 consists of models calculated with
CLES-013 (written by R. Scuflaire) with an initial mass fraction of metals
$Z_0$\,= 0.020 and of hydrogen $X_0$\,= 0.70, mixing-length $\alpha_{\rm
conv}$\,= 1.80, and the standard metal mixture of Grevesse \& Noels (1993).
Grid 2 was obtained with CLES-018.2 with
the 'new' solar values $Z_0$\,= 0.015, $X_0$\,= 0.71, $\alpha_{\rm conv}$\,=
1.75 and the standard metal mixture of Asplund et al. (2005).
In both cases, we used the CEFF equation of state (Christensen-Dalsgaard \&
D\"appen 1992) and a Kurucz atmosphere with the junction point at optical
depth $\tau$\,= 10, and we assumed neither convective overshooting nor
diffusion.
One of the main changes between CLES-013 and CLES-018.2 is the use of the new
value of the cross section of $^{14}N(p,\gamma)^{15}O$ recently measured by
Formicola et al. (2004).
We calculated the non-adiabatic eigenfunctions and eigenfrequencies for
$g$-modes with $\ell \leq 3$ with the code MAD (written by M.-A. Dupret).
For each star, we selected the models within the observed errorbox of
$\log(\teff)$ and \logg\ (boxes in Fig.\,\ref{hrmodel}), and selected
the eigenfrequency which is the closest to the observed frequency to calculate
the theoretical amplitude ratios.
In Fig.\,\ref{mifig}, we give a representative example of our results,
i.e. for the two main modes of HD\,179588.
Although there are significant differences in the position and/or the shape of
the theoretical curves for the higher degree modes of grid 1 (left) and 2
(right), the identification of the modes remains the same, i.e. $\ell$\,= 1 or
2 for the mode corresponding to \nua\,=  0.856543(15)~\cd\ (top), and
$\ell$\,= 1 for the mode corresponding to \nub\,= 2.04263(5)~\cd\ (bottom).
In general, these differences coming from the use of 2 different grids
increase for increasing values of the observed frequency. 
So far, all the well-identified SPB modes have $\ell$\,= 1 or 2.
For the $\beta$\,Ceps, the mode identification is still ongoing.

\section{Conclusions} \label{con}

Our photometric survey allowed a significant contribution in the
classification of variable B stars.
HD\,89688 is a possible $\beta$\,Cep/SPB hybrid star and the {\sc corot}
target HD\,180642 is a multiperiodic $\beta$\,Cep star of which the main mode
shows non-linear effects (Aerts et al., in preparation). 
Amongst the 31 targets with a sufficient amount of data, we identified 4
ellipsoidal variables and 4 spotted stars.  
Their classification should be checked by supplementary spectroscopic
observations. 
The mode identification is still ongoing, but all well-identified modes have
$\ell \le 2$ so far.

The final results of our survey will be given by De Cat et al. (in
preparation). 
The {\sc mercator} observations allow to take the first steps in asteroseismic
modeling for two multiperiodic $\beta$\,Cep stars, i.e. HD\,203664 (Aerts et
al., submitted to A\&A) and HD\,21803 (Saesen et al., in preparation). 
For 12\,Lac and V2052\,Oph, the {\sc mercator} telescope was included in
multi-site campaigns. 
The data of these objects are being analysed by Handler et al. (submitted to
MNRAS) and Handler et al. (in preparation) respectively. 

\acknowledgments{
This work is based on observations collected with the {\sc p7} photometer
attached to the {\sc mercator} telescope (La Palma, Spain).  
We are very much indebted to all the observers from Leuven University.
CA and JC acknowledge support from the Fund for Scientific Research (FWO) - Flanders
(Belgium) through project G.0178.02.   
}

\References{
Asplund M., Grevesse N., Sauval A.J., 2005, ASP Conf. Ser. 336, 25\\
Christensen-Dalsgaard J., D\"appen W., 1992, A\&ARv 4, 267\\
De Cat, P., De Ridder, J., Uytterhoeven, K., et al., 2004, ASP Conf. Proc. 310, 238\\
De Ridder, J., Cuypers, J., De Cat, P., et al., 2004, ASP Conf. Proc. 310, 263\\
Dupret M.-A., De Ridder J., De Cat P., et al., 2003, A\&A 398, 677\\
Formicola A., Imbriani G., Costantini H., et al., 2004, Physics Letters B 591, 61\\
Grevesse N., Noels A., 1993, Physica Scripta 47, 133\\
Pamyatnykh, A.A., 1999, Acta Astr. 49, 119\\
Scargle, J.D. 1982, ApJ 263, 835\\
Stellingwerf, R.F. 1978, ApJ 224, 953\\
}

\end{document}